\documentclass[aps,physrev,preprint,groupedaddress]{revtex4-2}

\usepackage{graphicx}
\usepackage{subcaption}
\usepackage{multirow}
\usepackage{array}
\usepackage{amsmath}
\usepackage{dcolumn}
\usepackage{bm}
\usepackage[bookmarks=false]{hyperref}

\begin{document}


\title{\textbf{Quantum Internet: Resource Estimation for Entanglement Routing} 
}%

\author{Manik Dawar}
\affiliation{Airbus Central Research and Technology, 82024 Ottobrunn, Germany
}%
\affiliation{
 Institute for Quantum Physics and Center for Optical Quantum Technologies, University of Hamburg,
Luruper Chaussee 149, 22761 Hamburg, Germany
}%

\author{Ralf Riedinger}
\affiliation{
 Institute for Quantum Physics and Center for Optical Quantum Technologies, University of Hamburg,
Luruper Chaussee 149, 22761 Hamburg, Germany
}%
\affiliation{
 The Hamburg Centre for Ultrafast Imaging, Hamburg, Germany
}%

\author{Nilesh Vyas}
\affiliation{%
 Airbus Central Research and Technology, 82024 Ottobrunn, Germany
}%

\author{Paulo Mendes}
\affiliation{%
 Airbus Central Research and Technology, 82024 Ottobrunn, Germany
}%

\date{\today}

\begin{abstract}
Quantum repeaters have promised efficient scaling of quantum networks for over two decades. Despite numerous platforms proclaiming functional repeaters, the realization of large-scale networks remains elusive,
indicating that the resources required to do so were thus far underestimated. 
Here, we investigate the dependence of resource scaling of networks on realistic experimental errors. Using a nested repeater protocol based on the purification protocol by Bennett et al., we provide an analytical approximation of the polynomial degree of the resources consumed by entanglement routing.
Our error model predicts substantially stricter thresholds for efficient network operation than previously suggested,
requiring two-qubit gate errors below 1.3\% for resource scaling with polynomial degree below 10. 
The analytical model presented here provides insight into the reason why previous experimental implementations of quantum repeaters failed to scale efficiently and inform the development of truly scalable systems, highlighting the need for high-fidelity local two-qubit gates.
We employ our analytical approximation of the scaling exponent as a figure of merit to compare different platforms and find that trapped ions and color centers in diamond currently provide the best route towards large-scale networks. 
\end{abstract}

\maketitle

\newpage
\tableofcontents

\section{Introduction}
\label{sec:intro}

Quantum networks are expected to unlock numerous applications of quantum technologies, such as global quantum key distribution (QKD) \cite{bennett2014quantum, ekert1991quantum, bennett1992quantum}, distributed quantum computing \cite{buhrman2003distributed}, sensing \cite{zhang2021distributed}, and blind quantum computing \cite{arrighi2006blind}. 
These applications are enabled by sharing entanglement between the networks' end-users. Hence, the main task of the network, which consists of quantum processor nodes connected by optical links, is to distribute and route entanglement efficiently and reliably. 

Direct photonic distribution of entanglement between multiple nodes is prone to transmission losses, resulting in data rates that decrease exponentially with distance in the case of optical fibers \cite{pirandola2017fundamental}.
This could be fixed using first-generation quantum repeaters \cite{dur1999quantum}: intermediate network nodes that store, purify, and swap entanglement, promising a polynomial scaling of the entanglement resources needed. 
Storage and swapping enables an asynchronous operation of individual, shorter sections of the link, avoiding the aforementioned exponential channel loss. Purification ensures that the entanglement fidelity remains above a threshold for long-distance connections, despite the increased number of imperfect operations. 

Network performance can potentially be further enhanced by advanced entanglement routing schemes, 
employing error correction \cite{hartmann2007role}, one-way communication \cite{borregaard2020one}, memory efficient \cite{childress2006fault} or  memory-less architectures \cite{li2019experimental}. 
Due to the stricter requirements of these protocols, near-term quantum networks are expected to employ first-generation quantum repeaters, which are hence investigated here. 

Indeed, multiple experimental platforms have demonstrated key features of quantum repeaters, lending hope for experimental implementations of large-scale quantum networks in the near future
\cite{bhaskar2020experimental, kalb2017entanglement, reichle2006experimental, yuan2008experimental, chou2007functional, pompili2021realization}. However, networks with more than three nodes \cite{pompili2021realization, hermans2022qubit}
have thus far not materialized, raising the question of what is actually required to make quantum networks feasible. 

Here, we propose an analytical, non-recursive approach for estimating the resource scaling of quantum networks based on a realistic experimental errors. This can provide valuable insights into the development of quantum networks by establishing a relation between the system performance of a physical platform and the polynomial degree $\lambda$ of the resources: the number of entangled photon-pairs required to establish entanglement across a distance $D$. This polynomial degree can serve as a key performance indicator for physical realizations of quantum networks, guiding their development towards full-scale implementations, supplementing the common understanding that entanglement rates need to exceed decoherence rates in memory-based systems.

First, we provide an overview of first-generation quantum repeaters and their analysis in existing literature in Sec.~\ref{sec:related work}, and then introduce our error model in Sec.~\ref{sec:purification}. Thereafter, we describe the non-recursive estimation of the entanglement resource consumption of a quantum network, 
and provide an estimation of the maximum size of a network due to decoherence of the memories
in Sec.~\ref{sec:exponent}. 
Finally, we provide a conclusion, estimating and comparing resource scalings for potential quantum networks built out of various experimental platforms in Sec.~\ref{sec:lambda kpi}.

\section{Related work}
\label{sec:related work}

\begin{figure}
    \centering
    \includegraphics[width=0.7\linewidth]{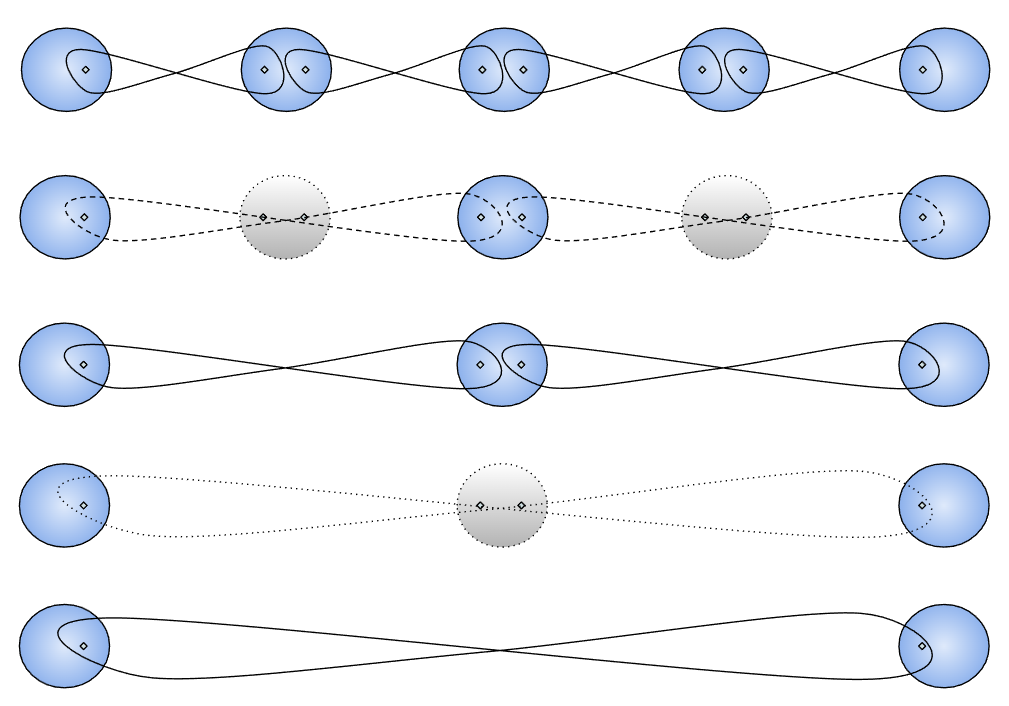}
    \caption{Nested repeater protocol, progressing from top to bottom. A linear repeater chain establishes high fidelity entanglement (solid lines) between neighbouring nodes (step 0). Swapping is performed at alternate nodes, thereby doubling the length of each connection, but introducing errors, resulting in lower fidelity (dotted lines). Therefore, in the next step (step 1), purification (Sec.~\ref{sec:purification}) is performed, using multiple long-distance, low fidelity entangled pairs to raise the fidelity back to the original level (solid lines in step 2). Hence, a self-similar repeater chain with longer distances is obtained. These nesting levels (steps 0 to 2) are repeated (step 3, 4), until a high fidelity link between the end nodes is established.}
    \label{fig:nested}
\end{figure}

First generation quantum repeaters \cite{briegel1998quantum, muralidharan2016optimal} route entanglement of a specific minimum fidelity along an arbitrary path in a quantum bipartite entanglement network through a sequence of entanglement swapping
\cite{briegel1998quantum} and purification \cite{bennett1992quantum, deutsch1996quantum} operations (see Fig.~\ref{fig:nested}).
Swapping increases the lengths of the direct links, while purification enhances their quality (entanglement fidelity). Both these operations consume entanglement resources:
Entanglement swapping employs a Bell state measurement between two quantum memories in a network node, called the quantum repeater, to combine two independently established adjacent links into one longer link, thereby consuming one link. As this inevitably introduces errors \cite{knill2007quantum}, purification is employed to counteract the resultant degradation in the entanglement fidelity. Purification combines multiple low-fidelity links into a single high-fidelity link. To achieve even longer distances, this process can be repeated, doubling the link lengths each time (nesting level), while maintaining the entanglement fidelity.

We note that it is possible to combine more than two links by entanglement swapping before purification. However, this is only beneficial in the limit of extremely low error rates, due to the concave nature of purification: the larger the error, the less efficient it becomes to purify the state back to its initial fidelity.
To analyze near-term networks, we hence restrict the protocol to a recursion of base two.

To establish entanglement over a distance of $D$ fundamental links, $L=\log_2D$ nesting levels are needed. If entanglement swapping introduces the same errors on each level, purification requires a constant overhead of $B$ low-fidelity links to obtain one high-fidelity link. Hence, it requires $2\cdot B$ high-fidelity links of the nesting level $k$ to create one high-fidelity link on level $k+1$, spanning twice the distance. By iteratively applying this procedure to all $L$ nesting levels, we find that one high-fidelity entangled qubit pair over distance $D$ requires 
\begin{equation}
    T = (2B)^{L} = D^\lambda
    \label{eq:polynomial_scaling}
\end{equation}
high-fidelity entanglement links of the lowest level, where the exponent, or polynomial degree, is
\begin{equation}
\lambda = \log_2B + 1.
\label{eq:original exponent}
\end{equation}
Hence, the average entanglement rate between two nodes separated by a physical distance $X$ in an equidistant network is approximated by:
\begin{equation}
R(X) = R_0\cdot \frac D T 
\sim \left(R_0 X_0^{\lambda-1}\right)\cdot {X}^{-\lambda+1},
\label{eq:entanglement rate}
\end{equation}
where $X_0$ and $R_0$ are the distance and entanglement rate of neighboring nodes, assuming all nodes operating in parallel. 
We note that an exact solution for analytically computing the exact probability distribution for the rate, or even its average, in quantum networks with imperfect storage and operations, is still an open problem \cite{bernardes2011rate, shchukin2019waiting, brand2020efficient, coopmans2022improved}.
Hence, interface bandwidth and the efficiency of photon coupling, frequency conversion and the entanglement protocol all contribute to a linear prefactor $\left(R_0 X_0^{\lambda-1}\right)$, whereas the rate-distance scaling inherits the exponent from \eqref{eq:polynomial_scaling}. We note that requiring constant $B$ on all nesting levels implicitly assumes negligible decoherence of the memory qubits \cite{hartmann2007role}, thus restricting the polynomial scaling to times $t\ll T_2$ much shorter than the memory coherence time $T_2$. Effectively, this results in the first condition for network implementations, $R_0 T_2\gg 1$, which can be fulfilled by a few platforms.

The second condition is a finite scaling exponent $\lambda < \infty$, with practical multi-node operation likely requiring $\lambda\ll 10$.
However, thus far it has been difficult to compare the scaling exponent of different platforms, as there has been no platform-independent analytical discussion thereof. 
While \eqref{eq:original exponent} looks simple, the number of low-fidelity links $B$ required by the purification step critically depends on employed protocol, operational errors, starting and target fidelity, etc. 
Furthermore, the foundations of quantum repeater theory predates the establishment of modern quantum information error treatment, and employ outdated error models \cite{briegel1998quantum, hartmann2007role, childress2006fault}, such that even previous numerical estimations of $\lambda$ may not be reliable.

Here, we focus on first-generation CNOT-gate based purification protocols \cite{bennett1996purification, deutsch1996quantum}. 
Other protocols \cite{hartmann2007role, childress2006fault} can further improve network performance, but require substantially higher performance or larger networks before this becomes relevant. 
The CNOT-based schemes involve two pairs of qubits, \(A_1B_1\) and \(A_2B_2\), which approximate the same entangled state, e.g.\ $\left|\Phi^+\right\rangle\propto\left|00\right\rangle+\left|11\right\rangle$ with fidelity \(F\). CNOT gates are applied between \(A_1 A_2\) and \(B_1 B_2\) respectively, and control bits \(A_1B_1\) are kept, if measurements of the target bits \(A_2\) and \(B_2\) in the computational basis yield the same results, and are discarded otherwise. Successful purification reduces bit-flip errors in the state, yielding a higher fidelity state. This is shown diagramatically in Appendix \ref{sec:bennett}, Fig.~\ref{fig:bennett purification}.
In the protocol by Bennett et.~al.~\cite{bennett1996purification}, 
the undesired components of the state ($\left|\Phi^-\right\rangle$, $\left|\Psi^-\right\rangle$, $\left|\Psi^+\right\rangle$) are randomized by local gates, resulting in a Werner state. 
This simplifies the analytical treatment, yielding a fidelity \(F'\) of \(A_1B_1\):
\begin{equation}
    F' = \frac{F^2 + \frac{(1 - F)^2}{9}}{F^2 + 2F\frac{(1 - F)}{3} + 5\frac{(1 - F)^2}{9}}
    \label{eq:purification ideal}
\end{equation}
after purification, neglecting local errors. 

The DEJMPS protocol \cite{deutsch1996quantum} does not discard the information in the distribution of the undesired states, but uses local rotations to alternate which components are suppressed. This results in a faster improvement in fidelity following the second round of purification, but does not allow for a recursive treatment of the nesting layers. Hence, to allow for an analytic treatment, we use the Bennett protocol here, and leave an analysis of the DEJMPS protocol for future work. 

\begin{figure}
\centering
  \centering
  \includegraphics[width=0.8\textwidth]{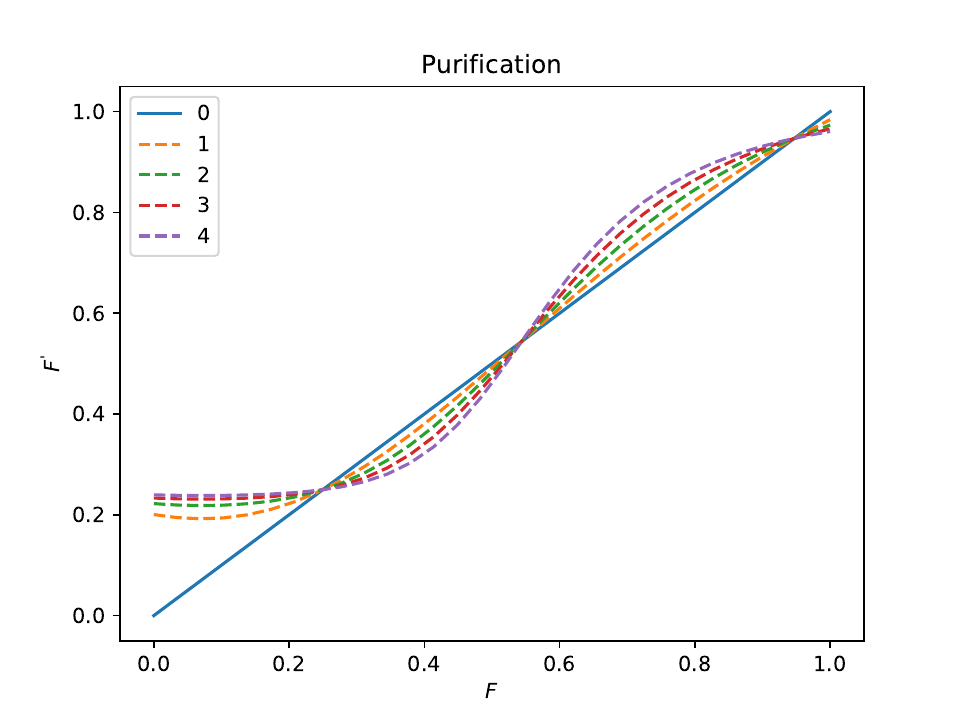}
\caption{Purification: Fidelity \(F'\) obtained on purifying two entangled pairs with fidelities \(F\) \cite{bennett1996purification} for exemplary gate and read error $\epsilon_g=0.01$, $\epsilon_R=0.01$.  The dashed curves represent successive applications (orange: once, purple: 4 times) of the purification map \eqref{eq:purification ideal}, while the solid blue line is the reference line: $F' = F$. In the regions where the dashed lines fall below the reference line, purification is not useful anymore. Hence, $F$ should be between the last two intersections between the dashed and solid lines.}
    \label{fig:purification ideal curve}
\end{figure}

The purification map \eqref{eq:purification ideal} is plotted in Fig.~\ref{fig:purification ideal curve}. It shows that purification is only possible between certain fixed points, in this case, 0.5 and 1. However, in the presence of errors, the fixed points gravitate towards each other, increasing the required minimum and reducing the maximum achievable fidelity \cite{briegel1998quantum}. 

A realistic analysis of quantum networks requires including local gate and readout errors. Briegel et.~al.~\cite{briegel1998quantum, dur1999quantum} considered an error model, in which the CNOT gate (in the purification protocol) performs ideally with probability $p_2$, but otherwise yields a completely mixed state. This model was subsequently used to analyze different repeater protocols  \cite{childress2006fault, hartmann2007role}.
However, it creates an abundance of correlated errors, i.e.\ in half of the instances, a gate error will affect the target qubit, resulting in anti-correlated measurements of $A_2$ and $B_2$, and thus dismissal of the state. Physical qubits can have vastly different error rates on source and target qubits. A single qubit error process affecting only the source qubit ($A_1$ or $B_1$) goes unnoticed by CNOT-based purification schemes, thus strongly degrading their performance.  

Here, we propose an error model that can be applied to CNOT-based purification protocols \cite{dur2007entanglement}. The model offers the flexibility of considering different probabilities for different Pauli errors. This is motivated by the observation that gate errors can be well described by single qubit errors occuring before or after an ideal gate operation \cite{song2019quantum},
as well as that qubits can have vastly different relaxation and decoherence times, resulting in substantially different flip and phase error probabilities \cite{wang2021single, chatterjee2021semiconductor, place2021new}. 

\section{Error model: purification and swapping}
\label{sec:purification} 
Given two entangled states with fidelities $F$, that are subject to the purification process, we consider two sources of experimental errors: the read-out efficiency $\eta'$ and gate errors on the control and target qubits, $\epsilon_s$ and $\epsilon_t$. 
Specifically, an imperfect measurement on a single qubit can be modeled as \cite{dur1999quantum}: 
\begin{equation}
\begin{aligned}
    P_0^{\eta'} &= \eta' |0\rangle\langle0| + (1-\eta')|1\rangle\langle1| \\
    P_1^{\eta'} &= \eta' |1\rangle\langle1| + (1-\eta')|0\rangle\langle0|
\end{aligned}
\label{Eq:measure}
\end{equation}
Therefore, if the qubit to be measured is $\rho = |0\rangle\langle0|$, then the expectation values are $\langle P_0^{\eta'} \rangle = \eta'$ and $\langle P_1^{\eta'} \rangle = 1-\eta'$, meaning that there is a probability $\eta'$ that the measurement (\ref{Eq:measure}) gives the correct result $|0 \rangle \langle 0|$. 
This could be generalized to asymmetric read errors, see Appendix \ref{sec:bennett}.

We model a noisy CNOT gate as an ideal gate, which is occasionally accompanied by a Pauli error, i.e.\ with some probability $p_{i,j,k}$, a Pauli operator $\sigma_j^i$ ($j=x,y,z$) acts on the source ($i=s$) or target qubit ($i=t$), either before ($k=b$) or after ($k=a$) the ideal gate (see Appendix \ref{sec:bennett}). While such a model can reproduce experimental observations reasonably well \cite{song2019quantum}, its description requires 13 parameters per node, which are rarely fully characterized in state-of-the-art experimental platforms. For the sake of simplicity, we group subsequent errors in a multi-round purification protocol, allowing for a reduction of the complex model to an effective gate error $\epsilon_g$, and an effective read-out error $\epsilon_r=1-\eta'$: 

In other words, the effective gate error encompasses errors happening in between two purification rounds, i.e.\
any error affecting the source qubit after the $n$-th, and before the $(n+1)$-th idealized CNOT gate. Effectively, this means that a purification step $F\rightarrow F'$ describes the change in the conditional state fidelity available to the next idealized entanglement gate. This is useful, when the distributed entanglement will be used e.g.\ in quantum communication (for teleportation), or to route entanglement (via entanglement swapping). Thus the state $\rho$ immediately before the idealized CNOT operation $U$ transforms to: 
\begin{equation}
    \rho' \approx (1-\epsilon_g)U \rho U^\dagger + \epsilon_g \sum_j p_j (\sigma_j^s [U \rho  U^\dagger] \sigma_j^{s\dagger}),
    \label{eq:error model_after}
\end{equation}
where $\epsilon_g$ is the probability of an error effecting the source qubit, and $p_j$ the conditional probability of this error being the Pauli operator $\sigma_j^s$ acting on the source qubit ($j=x,y,z$).

Similarly the effective read-out error contains the physical readout error, as well as bit flip errors on the target qubit after the idealized gate. 
\begin{equation}
    \eta:=(1-p_{t,x,a})(1-p_{t,y,a})\eta'.
\end{equation}
Applying this error model to the purification map (\ref{eq:purification ideal}) and neglecting terms that are quadratic or higher order in the errors $1-\eta$ and $\epsilon_g$ (such as independent, simultaneous read-out and gate errors), we get, the fidelity $F'$ after a successful purification step (see Appendix \ref{sec:purification map} for the derivation):
\begin{equation}
\begin{matrix}
    F'(F,\eta, \epsilon_g) = \frac{\left[ F^2 + \left(\frac{1-F}{3}\right)^2 \right] \left[\eta^2 + (1 - \eta)^2\right] + \left[F\left(\frac{1-F}{3}\right) + \left(\frac{1-F}{3}\right)^2\right]2\eta(1 - \eta) + 2\left(\frac{2\epsilon_g - \epsilon_g^2}{(1-\epsilon_g)^2}\right)\left[p_z F\left(\frac{1-F}{3}\right) + (p_x + p_y) \left(\frac{1-F}{3}\right)^2\right]}{\frac{1}{(1-\epsilon_g)^2}\left( \left[ F^2 + 2F\left(\frac{1 - F}{3}\right) + \frac{5}{9}\left(1 - F\right)^2 \right] \left[\eta^2 + (1 - \eta)^2 \right] + \left[F\left(\frac{1-F}{3}\right) + \left(\frac{1-F}{3}\right)^2\right]8\eta(1 - \eta) \right)}
\label{eq:purification eta eg}
\end{matrix}
\end{equation}
Specifically, the probability $P_F$ of the purification operation being successful is proportional to the denominator:
\begin{equation}
\begin{matrix}
    P_F =  \left[ F^2 + 2F\left(\frac{1 - F}{3}\right) + \frac{5}{9}\left(1 - F\right)^2 \right] \left[\eta^2 + (1 - \eta)^2 \right] + \left[F\left(\frac{1-F}{3}\right) + \left(\frac{1-F}{3}\right)^2\right]8\eta(1 - \eta)
\end{matrix}
    \label{eq:pf}
\end{equation}
In the rest of the paper, we take $p_z = p_x + p_y = 1/2$ in (\ref{eq:purification eta eg}) for simplicity. 
We note that this expression is similar to the fidelity map used in \cite{dur1999quantum}. However, here $P_F$ is independent of the effective gate error. This makes sense because it only affects the source qubit. 

The error model (\ref{eq:error model_after}) can also be applied to entanglement swapping, which joins $N$ adjacent links of fidelity $F$ into a longer one with fidelity $F_N\leq F$ \cite{dur1999quantum}: 

\begin{equation}
    F_N = \frac{1}{4}\left\{1 + 3[(1-\epsilon_g)^3]^{N-1} \left(\frac{4 \eta^2 - 1}{3}\right)^{N-1} \left(\frac{4F-1}{3}\right)^N \right\}
    \label{Eq:fidelity us without absorbing}
\end{equation}
However, to be consistent with our own convention from the purification map \eqref{eq:purification eta eg}, where we "absorb" the gate error on the target qubit into the read-out efficiency, we do the same with swapping, but since both qubits are measured during a Bell State Measurment, we absorb the gate errors into the read-out efficiencies $\eta_{s}$ on both the qubits:
\begin{equation}
    F_N = \frac{1}{4}\left\{1 + 3 \left(\frac{4 \eta_s^2 - 1}{3}\right)^{N-1} \left(\frac{4F-1}{3}\right)^N \right\}
    \label{Eq:fidelity us}
\end{equation}
We assume a further simplification in the proceeding analysis by taking the effective read-out efficiency $\eta_s$ in swapping to be the same as that on the target qubit in purification: $\eta_s=\eta$. This is a reasonable assumption, since swapping and purification are expected to be performed on the same hardware. 

\section{Non-recursive estimation of resources}
\label{sec:exponent}
\subsection{Nested repeater protocol}
\label{sec:nested protocol}

We illustrate our error model (Sec.~\ref{sec:purification}) on a nested repeater protocol (Sec.~\ref{sec:nested protocol}) using the Bennett purification protocol \cite{bennett1996purification}, in which we assume error-free depolarising steps. We suppose that we start with a linear chain, with entanglement of fidelity $F_t$ between neighbouring nodes. On swapping, the fidelity drops to $F_0$, given in terms of $F_t$ by substituting $N=2$ and $F = F_t$ in the swap-fidelity equation (\ref{Eq:fidelity us}). We desire to recover the original fidelity $F_t$ through repeated purification. Realistically, since the link length grows with each nesting level, so does the latency in the classical communication between the connected nodes. This should result in a worse $F_0$ at each nesting level due to decoherence. We neglect this effect for simplicity and assume the same $F_0$ and $F_t$ at each nesting level. 

\subsection{Resource Estimation}
Given swap and purification success probabilities $P_s$ and $P_{F_i}$, the number of entanglement pairs $B$ required in each nesting level:
\begin{equation}
    B = \frac{2^m}{P_s^m \prod_{i=1}^m P_{F_i}}
    \label{eq:B}
\end{equation}
Here $m$ is the number of successful purification operations required to go from a diminished fidelity $F_0$ to a target fidelity $F_t$. To the best of our knowledge, no non-recursive expressions exist for $B$.

There are two aspects of (\ref{eq:B}) that make a non-recursive estimation of $B$ challenging. The number of purification steps in each nesting level, $m$, itself cannot be exactly expressed non-recursively. Moreover, each $P_{F_i}$ in the product in the denominator depends on $F_i$ and hence changes with every step from 1 to $m$. Consequently, the product $\prod_{i=1}^m P_{F_i}$ cannot be expressed non-recursively either. 

We approach a non-recursive approximation $\Tilde{m}$ of the required number of successful purification steps $m$ by computing the average improvement size $g$ in the region bounded by the start and target fidelities $F_0$ and $F_t$:
\begin{equation}
    g = \frac{\int_{F_0}^{F_t} [F'(F) - F] dF}{F_t-F_0}
    \label{eq:approx g}
\end{equation}
Where $F'(F)$ is given by (\ref{eq:purification eta eg}). Hence, 
\begin{equation}
\begin{split}
\Tilde{m} &= \left\lceil \frac{\Delta F}{g} \right\rceil
\end{split}
\label{eq:approx m}
\end{equation}
Where $\Delta F := F_t-F_0$. For the exact expression of $\Tilde{m}$, we refer the reader to Appendix \ref{sec:full integrals}: (\ref{eq:m numer}) and (\ref{eq:m denom}). 

The product of the purification acceptance probabilities $\prod_{i=1}^m P_{F_i}$ in the denominator of (\ref{eq:B}) is approximated by computing the geometric mean of the acceptance probabilities $\Tilde{P_F}$ in the improvement interval $[F_0,F_t]$. This can be computed using a geometric integral of $P_F$ over the interval \cite{slavik2007product}:
\begin{equation}
    \Tilde{P_F} = \lim_{m \rightarrow \infty} \left( \prod_{i = 1}^m P_{F_i} \right)^{1/m} = \exp\left(\frac{\int_{F_0}^{F_t} \log P_FdF}{F_t-F_0}\right)
    \label{eq:avg pf}
\end{equation}
The explicit expression for the integral in (\ref{eq:avg pf}) is also given in Appendix \ref{sec:full integrals}: (\ref{eq:pf explicit}).

\begin{figure}
    \centering
    \includegraphics[width=0.7\textwidth]{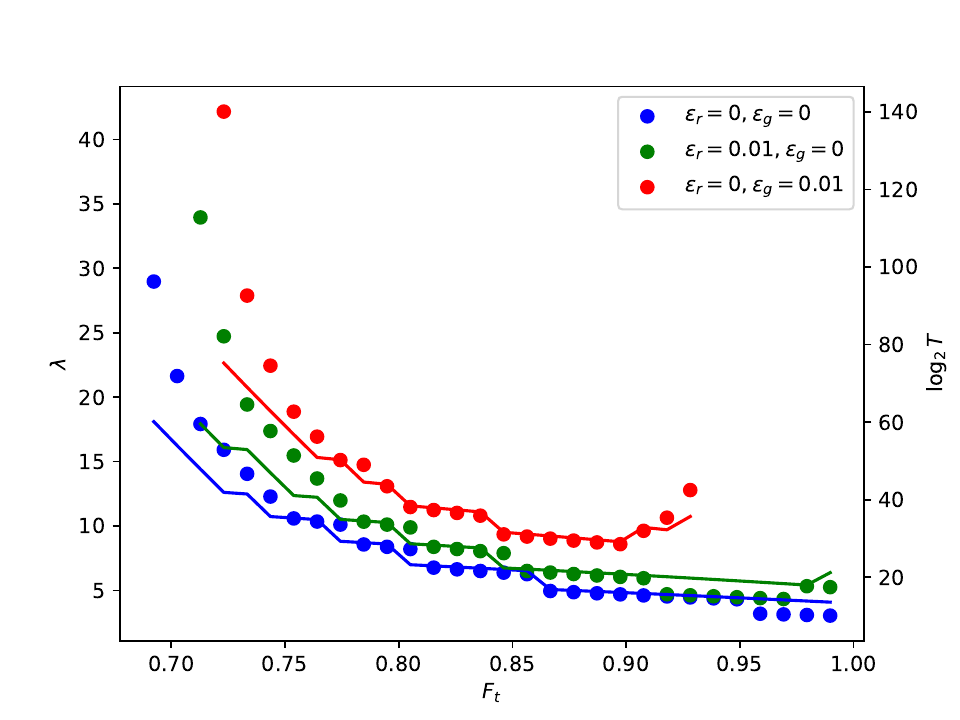}
    \caption{Exponent $\lambda$ (left $y$-axis) computed recursively from (\ref{eq:purification eta eg}) (dots) and non-recursively (solid lines) from (\ref{eq:approx exp}) for different read-out and gate errors: $\epsilon_r$ and $\epsilon_g$. The $y$-axis on the right is $\log_2$ of the total resources $T$ needed to route entanglement along a path of 10 hops. Here, by definition, $T=D^\lambda$} 
    \label{fig:recursive vs non} 
\end{figure}
Where $\alpha \in \{0,1\}$ and $F_1:=F_t$. 
The approximation for the total number of entanglement pairs $B$ (\ref{eq:B}) needed to purify from $F_0$ to $F_t$ is thereby complete: 
\begin{equation}
    \Tilde{B} = \left( \frac{2}{P_s \Tilde{P_F}} \right)^{\Tilde{m}}
    \label{eq:approx B}
\end{equation}
For the rest of the discussion, we take swapping to be a deterministic operation: $P_s=1$. Now an approximation of the exponent $\lambda$ from \eqref{eq:polynomial_scaling} can be expressed non-recursively as:
\begin{equation}
    \Tilde{\lambda} = \log_2 \Tilde{B} + 1
    \label{eq:approx exp}
\end{equation}
See Appendix \ref{sec:full integrals} for the full expansion of all the terms in (\ref{eq:approx exp}). 

In contrast to the exact exponent in (\ref{eq:original exponent}), the computation of which, requires $m$ (recursion) steps, the approximation (\ref{eq:approx exp}) is just a single expression that can be computed at once, bringing down the time-complexity of the computation from linear to constant in $m$. The potential errors in these approximations: (\ref{eq:approx m})  and (\ref{eq:avg pf}), arise from the fact that in an actual experiment, there is a finite number of improvement steps, but an infinite number of them is required in general to reproduce the true average in the improvement interval $[F_0,F_t]$. Therefore, in the limit $m\rightarrow\infty$, these approximations becomes exact. 

In Fig.~\ref{fig:recursive vs non}, we compare the exponents, as computed recursively from (\ref{eq:purification eta eg}) with those computed non-recursively from (\ref{eq:approx exp}), for different read-out and gate errors: $\epsilon_r:=1-\eta$ and $\epsilon_g$. Here, we obtained $F_0$ by substituting $F$ with $F_t$, and $N$ with 2 in (\ref{Eq:fidelity us}). It seems from (\ref{fig:recursive vs non}), that for each error combination, there is generally an optimal target fidelity $F_t$ which minimises the resource consumption $\lambda$ of a quantum network. This can be reasoned as follows: \\The probability $P_F$ of a successful purification operation (\ref{eq:pf}) gets better as the fidelity $F$ gets better, contributing to a lower $\lambda$. On the other hand, as we can see from Fig.~\ref{fig:purification ideal curve}, the improvements in fidelity get smaller at larger fidelities, contributing to a larger $\lambda$. These trade-offs suggest the existence of an optimal fidelity $F_t^*$ that one should target, given gate and read-out errors, that minimizes $\lambda$. 

\begin{figure}[h!]
\centering
    \includegraphics[width=0.8\textwidth]{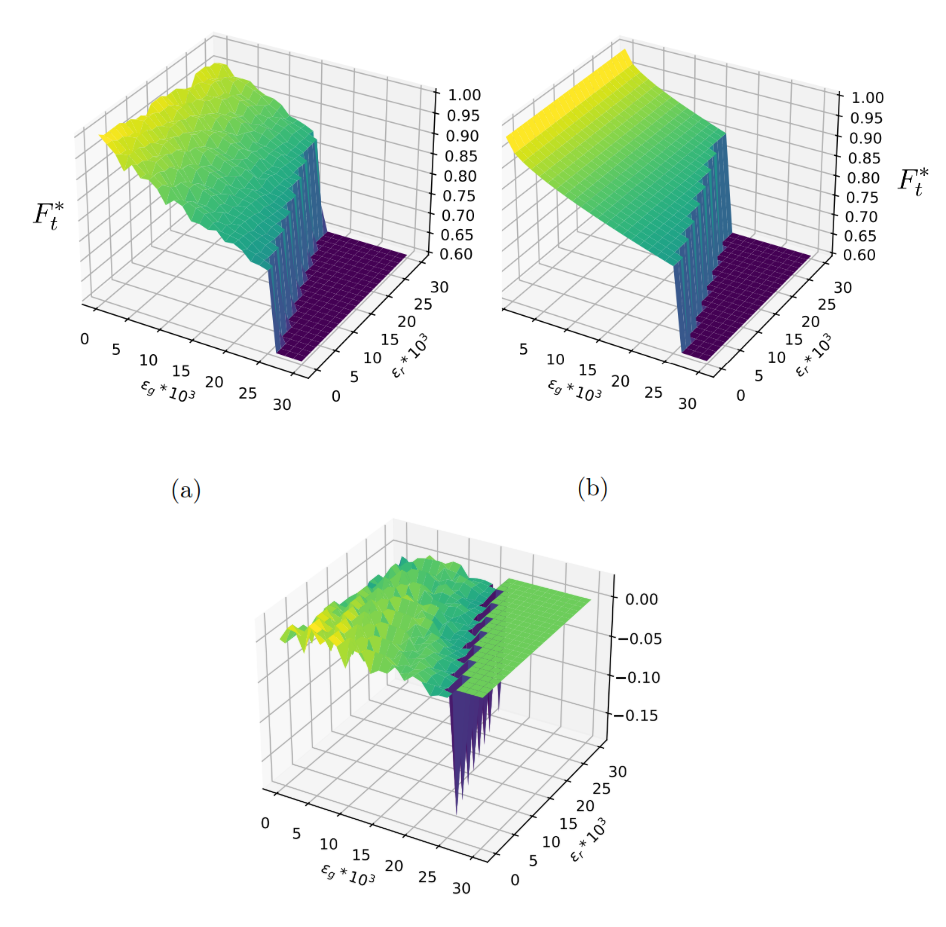}
\caption{3D surface plots for the optimal target fidelity $F_t^*$ that minimises the resource scaling $\lambda$. Fig.~(a) plots $F_t^*$ as obtained from recursive applications of (\ref{eq:purification eta eg}), and Fig.~(b) does this using the simple relation in (\ref{eq:opt ft}). Fig.~(c) takes the difference between the simulations: Fig.~(a) and the analytical approximation: Fig.~(b). In the error regions in (a) and (b) where purification is infeasible, we just take $F_t^*$ to be 0.6 for illustration purposes.}
\label{fig:heatplots ft}
\end{figure}

Indeed, we obtained an analytical approximation of this optimal fidelity $F_t^*$ by minimising an approximation of $\lambda$, with respect to the target fidelity $F_t$: 
\begin{equation}
    F_t^* \approx \frac{-1.16\epsilon_g - 4.28\sqrt{\epsilon_g^2 + 0.15\epsilon_g} + 1.9}{2.66\epsilon_g + 1.9}
    \label{eq:opt ft}
\end{equation}
Notably, we approximated $F_0$, as obtained from (\ref{Eq:fidelity us}), with $2F_t-1$. Moreover, we observed very little dependence of $F_t^*$ on $\epsilon_r$, so we neglected it. For a more detailed derivation of (\ref{eq:opt ft}), see Appendix \ref{sec:opt ft}.  
Fig.~\ref{fig:heatplots ft} shows a comparison of the optimal target fidelity $F_t^*$ obtained by numerically optimizing \eqref{eq:approx B}, and the approximation \eqref{eq:opt ft}. We find, in accordance with our approximation, that $F_t^*$ primarily depends on the gate error $\epsilon_g$, with the readout error $\epsilon_r$ only noticeably influencing the maximal gate-error that allows for purification. 
This critical error happens when swapping entanglement with a fidelity corresponding to the larger fixed point results in a fidelity $F_0$ (see eq. \ref{Eq:fidelity us}) below the lower fixed point. Note that in these error regions where purification is infeasible, we just take $F_t^*=0.6$ for illustration purposes.  
These fixed points were computed exactly from \eqref{eq:purification eta eg} and used in the 3D-plots to determine the transition points, which are not directly apparent in \eqref{eq:opt ft}. 
Fig. \ref{fig:heatplots ft} shows that the approximation agrees with simulations for the most part, except for systematic deviation close to the critical errors,
where purification goes from being feasible to infeasible. For a derivation of (\ref{eq:opt ft}), see Appendix \ref{sec:opt ft}.

\begin{figure}[ht!]
\centering
    \includegraphics[width=0.8\textwidth]{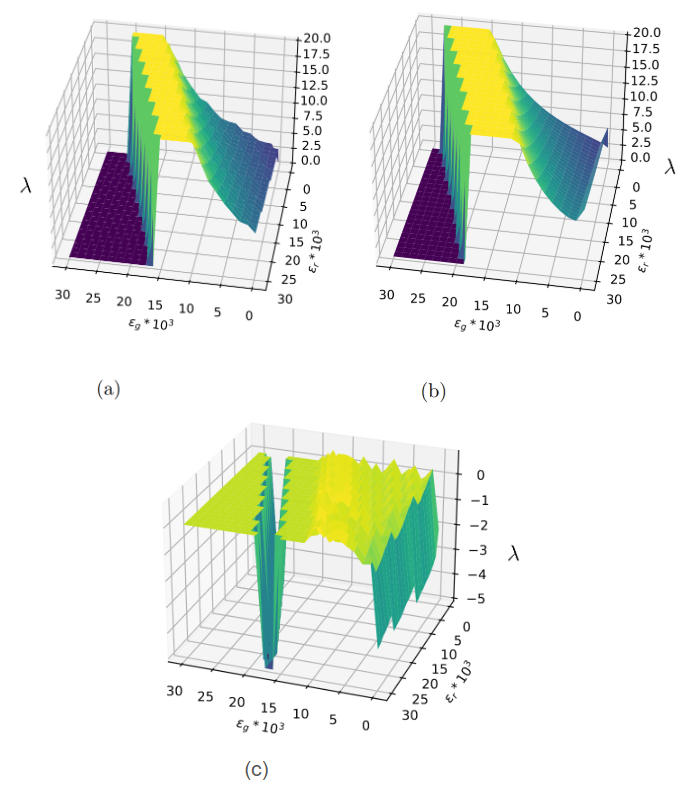}
\caption{3D surface plots for the optimal resource scaling exponent $\lambda$ in terms of the network errors $\epsilon_r$ and $\epsilon_g$. The optimal $\lambda$ is obtained by deciding $F_t$ using (\ref{eq:opt ft}), and $F_0$ from $F_t$ using (\ref{Eq:fidelity us}). Fig.~(a) is obtained from repeated applications of (\ref{eq:purification eta eg}) until $F_T$ is reached or exceeded. Fig.~(b) follows from our non-recursive expression (\ref{eq:approx m}). We find an exponent $\lambda<10$ requires a gate error below $\epsilon_g<0.013$. In Fig.~(c), we plot the difference between Figs.~(a) and (b).}
\label{fig:heatplots}
\end{figure}

Using \eqref{eq:approx exp} and \eqref{eq:opt ft}, we obtain an explicit, non-recursive estimation for the optimal exponent $\Tilde\lambda$. This expression fairly large, and does not yet provide a detailed insight. For small gate error $\epsilon_g$, near perfect starting fidelity $F$, and neglecting readout error $\epsilon_r$ (due to the small dependence, and routinely achieved experimental values, see Fig.~\ref{fig:heatplots} and Table \ref{tab:lambdas}), however, this can be approximated by:  
\begin{equation}
    \Tilde{\lambda} \approx \Tilde{\lambda}_s = 3 + 14 \sqrt{\epsilon_g} + 38 \epsilon_g
    \label{eq:approx exp_small_errors}
\end{equation}
We find a strong dependence of the exponent on the effective gate error, which further increases for larger errors. 
The limit of $\lambda\geq 3$ corresponds to a minimum of 2 iterations of purification, in the presence of non-zero errors, required by the Bennett purification protocol to recover the original, near-ideal fidelity $F_T$. The limit of $\Tilde{\lambda}_S \geq 3$ is not fundamental, however, and can be improved upon, given low gate errors, by performing entanglement swapping over multiple ($n$) links, before purifying the state. For a base-$n$-protocol, the limit becomes $\Tilde{\lambda}_S \geq 2\log_n 2+1$, and will be topic of future studies. 

Fig.~\ref{fig:heatplots} compares the recursivly and non-recursivly estimated exponents: (\ref{eq:original exponent}) and (\ref{eq:approx exp}) with respect to the read-out and gate errors $\epsilon_r:=1-\eta$ (\ref{Eq:measure})  and $\epsilon_g$ (\ref{eq:error model_after}). In Fig.~\ref{fig:heatplots}a, we compute $\lambda$ through recursive applications of the swapping and purification maps \eqref{Eq:fidelity us} and (\ref{eq:purification eta eg}) at the optimal target fidelity $F_t^*$ for the given error configuration ($\epsilon_r, \epsilon_g$), as shown in Fig.~\ref{fig:heatplots ft}a. In Fig.~\ref{fig:heatplots}b, we plot the non-recursive exponent $\Tilde{\lambda}$ (\ref{eq:approx exp}), without the ceiling function in (\ref{eq:approx m}), at target fidelity $F_t^*$ from Fig.~\ref{fig:heatplots ft}b. In both cases, $F_0$ was computed from from $F_t^*$ using (\ref{Eq:fidelity us}), resulting in the errors ($\epsilon_r, \epsilon_g$) being the only free parameters. 

Beyond the threshold errors $\epsilon_g$ and $\epsilon_r$, the exponent $\Tilde{\lambda}$ becomes negative, indicating that purification becomes infeasible in those error regimes. For illustration purposes, these negative values are set to zero, to allow for 
a clear visual distinction between the feasible and the infeasible error regimes: the hill on the right, and the valley on the left respectively. 
For the recursive 3D-plot in Fig.~\ref{fig:heatplots}a, the infeasible error regimes were identified by computing analytically, the last two fixed points ($F_l$ and $F_u$) of the purification map (\ref{eq:purification eta eg}); then if either $F_0$ or $F_t$ fell outside the open interval $[F_l, F_u]$, that implied the error combination $(\epsilon_r, \epsilon_g)$ was infeasible for purification. These infeasible regions are likewise also shown as 0's in the plot. Note that the maximum value shown in the plots is capped at 20 for clarity. Anyway, we suspect, any network with an exponent larger than that would be hugely impractical. 

In Fig.~\ref{fig:heatplots}c, we plot the difference between the numerical data in Fig.~\ref{fig:heatplots}a and the analytical data in Fig.~\ref{fig:heatplots}b. The absolute difference is small in most places, apart from the transition region, where the error regime transitions from feasible to infeasible and the $\epsilon_g = 0$ line. At the $\epsilon_g = 0$ line, this is because of the disparity in the simulated and analytical optimal $F_t^*$ (\ref{eq:opt ft}). 

\subsection{Maximum path-length}
Eq.~\eqref{eq:polynomial_scaling} shows us that the entanglement resources $T$ required for routing end-to-end entanglement grow polynomially with the length $D$ of the routing path as $D^\lambda$. Therefore, $T$ entangled qubits need to be distributed between each pair of neighboring nodes along the path. The qubits stored in the memories, however, suffer decoherence, which causes the fidelity $F$ of a stored entanglement to diminish with time $t$ as \cite{wang2016experimental}:
\begin{equation}
    F(t) = F(0)e^{-\left(\frac{1}{T_2} t\right)^2}, 
\end{equation}
where $T_2$ is the decoherence time under dynamical decoupling from the environment. This limits the neighbor-neighbor entanglements that could be set up and swapped before decoherence makes the fidelity drop below the lower fixed point from (\ref{eq:purification eta eg}). This condition is given by (\ref{eq:battle decoherence}):
\begin{equation}
    \frac{1}{4} \left( 1 + 3\left(\frac{4\eta^2 - 1}{3}\right) \left( 4F^*_t e^{-\left(\frac{1}{T_2}\frac{D^{\lambda}}{R}\right)^2} - 1\right)^2\right) > F_l
    \label{eq:battle decoherence}
\end{equation}
Where $R$ is the neighbor-neighbor entanglement generation rate and $F_l$ is the lower fixed point from (\ref{eq:purification eta eg}). We used the swapping fidelity (\ref{Eq:fidelity us}) in (\ref{eq:battle decoherence}) to account for the decrease in fidelity from swapping. An approximation here is that all stored entanglements have the same fidelity. This could be refined in future studies.

Setting the left and right side of \eqref{eq:battle decoherence} equal, we find a maximum feasible path-length (number of links between quantum repeaters) $D^*$ along which entanglement could be routed and maintained at the optimal target fidelity $F_t^*$:
\begin{equation}
     D^* = \left\lfloor \left( R T_2 \sqrt{-\log{\left(\frac{3 \sqrt{\frac{4 F_{l} - 1}{4 \eta^{2} - 1}} + 1}{4 F_t^*} \right)}}\right)^{\frac{1}{\lambda}} \right\rfloor
     \label{eq:chain length}
\end{equation}

\section{Discussion and conclusion}
\label{sec:lambda kpi}

We investigated the dependence of the resource scaling exponent $\lambda$ in a quantum network, based on gate and readout errors. This exponent can serve as a key performance indicator for experimental repeater implementations, supplementing entanglement rate and coherence time. For example, it can be employed to predict across how many links quantum information could be transferred before decoherence becomes relevant. 
To enable a more thorough understanding of the key performance indicator, we developed an analytical, non-recursive estimation of the exponent $\Tilde\lambda$.

\begin{table}
\centering
\caption{Resource exponent $\lambda$ for different quantum computing platforms. Each platform is characterized by a different gate and read-out error combination: $\epsilon_g$ and $\epsilon_r$. Using this, $\Tilde{\lambda}$ was computed using (\ref{eq:approx exp}) without the ceiling function in \eqref{eq:approx m} and $\lambda$, using recursive applications of (\ref{eq:purification eta eg}). Then, using the neighbor-neighbor entanglement generation rate $R$, and the decoherence time $T_2$ from recent experiments, the maximum number of hops $D^*$ was computed from \eqref{eq:chain length} without the floor function. 
}
\label{tab:lambdas}
\begin{tabular}{|c|c|l|l|l|l|l|l|}
\hline 
\textbf{Platform} & $\boldsymbol{\epsilon_g}$ & $\boldsymbol{\epsilon_r}$ & $\boldsymbol{\Tilde{\lambda}}$ & $\boldsymbol{\lambda}$ & $\boldsymbol{R}$ & $\boldsymbol{T_2(s)}$ & $\boldsymbol{D^*}$\\ \hline 
Superconducting & 2.5$\times 10^{-3}$ \cite{xu2020high}& 6$\times 10^{-3}$ \cite{jurcevic2021demonstration} & 5.1 &  5.82 & 5.4 \cite{riedinger2018remote} & 0.3 $\times 10^{-3}$ \cite{crowley2023disentangling} & 0.3 \\ \hline
SiV centers & $5 \times 10^{-4}$ \cite{stas2022robust} & $10^{-4}$ \cite{bhaskar2020experimental}& 3.49&4.06 & 1 \cite{knaut2024entanglement} & 2.1 \cite{stas2022robust} & 1.06 \\ \hline
NV centers & $3.5 \times 10^{-4}$ \cite{bartling2024universal} & 4 $\times 10^{-4}$ \cite{zhang2021high} & 3.47&4.11 & 39 \cite{humphreys2018deterministic} & 1 \cite{abobeih2018one} & 2.15 \\ \hline
Trapped ions & $5 \times 10^{-4}$ \cite{foss2024progress}& $10^{-6}$ \cite{erickson2022high}& 3.47 & 4.01 & 250 \cite{o2024fast} & 0.14 \footnote{Trapped ions may exhibit much longer coherence times \cite{wang2021single}, but this was the longest one we are aware of, in an optically connected setup.}\cite{drmota2023robust}& 2.14 \\ \hline
Neutral atoms &  2.5 $\times 10^{-3}$ \cite{evered2023high} & $4 \times 10^{-3}$ \cite{radnaev2024universal}&4.84 &5.62 & 0.11 \cite{ritter2012elementary} & $10^{-2}$ \cite{korber2018decoherence} & 0.27 \\ \hline  
\end{tabular}
\end{table}
In Table \ref{tab:lambdas}, we estimate the exponent $\lambda$ for various hardware platforms for quantum network nodes by considering state-of-the-art values for the gate error $\epsilon_g$ and read-out error $\epsilon_r$. 
We note that the error rates
were not taken from the same experimental setup, and they cannot necessarily be realized simultaneously in the same setup. Hence, this represents an optimistic estimation of what will be experimentally possible in the near future. 
Some additional assumptions that went into Table \ref{tab:lambdas} have been listed in Appendix \ref{sec: table assumptions}. The numerical estimation of the exponent $\lambda$ was computed through recursive applications of the purification map (\ref{eq:purification eta eg}), while $\Tilde{\lambda}$ comes from our analytical approximation (\ref{eq:approx exp}) without the ceiling function in \eqref{eq:approx m}. $R$ is the entanglement generation rate between two nodes of the corresponding platform. $D^*$ (\ref{eq:chain length}) is an estimate of the maximum length of a repeater chain constructed from nodes of the platform, beyond which, link fidelities cannot be maintained. Note that $D^*$ was computed from \eqref{eq:chain length} without the floor function.

For fiber-based long-distance quantum communication, conversion of itinerant photons to the telecom wavelength is required. Of the platforms investigated here, silicon vacancy color centers in diamond (SiV) offer the best performance, with all parameters demonstrated in the same setup. 

For local networks, such as quantum supercomputers, nitrogen vacancy centers in diamond (NV) and trapped ions offer the best performance, with ions standing to profit from potential improvement in the coherence time by sympathetic cooling. 

In summary, we investigated the dependence of resource scaling in quantum networks on realistic experimental errors. We provided an analytical approximation of the resource scaling exponent $\Tilde\lambda$, finding that gate errors are most critical for an efficient network operation. 
Our single-qubit Pauli error model results in a stricter threshold for the gate error $\epsilon_g< 2.88\%$ compared to historical error analysis, which suggested $\epsilon_g< 5\%$ was sufficient \cite{briegel1998quantum}. We find a strong dependency of the exponent on the gate errors, requiring $\epsilon_g\leq1.3\%$ for a distance scaling with an exponent $\tilde{\lambda}\leq 10$, which we choose as arbitrary reference value for experimental feasibility of quantum networks.
Overall, based on our analysis, quantum repeaters based on trapped ions and color centers in diamond offer the best resource scaling, potentially allowing for entanglement purification in multi-node networks. In summary, while there remain some technical challenges to be addressed, state-of-the-art quantum processors could enable polynomial resource scaling with an exponent $\lambda<4$ in the near future. 

\section*{Acknowledgment}
This research was co-financed by Airbus. 
 RR acknowledges support from the Cluster of Excellence ‘Advanced Imaging
of Matter’ of the Deutsche Forschungsgemeinschaft (DFG)–EXC 2056–Project ID 390715994, Bundesministerium für Bildung und Forschung (BMBF) via project QuantumHiFi - 16KIS1592K - Forschung Agil. The project is co-financed by ERDF of the European Union and by 'Fonds of the Hamburg Ministry of Science, Research, Equalities and Districts (BWFGB)'.

\appendix

\section{Purification map by Dur et.~al.}
\label{sec: briegel purification}
The purification map from Dur et.~al.~\cite{dur1999quantum} is given by:
    \begin{equation}
    \begin{matrix}
    F'(F,\eta, p_2) = \frac{\left[ F^2 + \left(\frac{1-F}{3}\right)^2 \right] \left[\eta^2 + (1 - \eta)^2\right] + \left[F\left(\frac{1-F}{3}\right) + \left(\frac{1-F}{3}\right)^2\right]2\eta(1 - \eta) + \frac{1-p_2^2}{8p_2^2}}{\left[ F^2 + 2F\left(\frac{1 - F}{3}\right) + \frac{5}{9}\left(1 - F\right)^2 \right] \left[\eta^2 + (1 - \eta)^2\right] + \left[F\left(\frac{1-F}{3}\right) + \left(\frac{1-F}{3}\right)^2\right]8\eta(1 - \eta) + \frac{1-p_2^2}{2p_2^2}}
    \end{matrix}
\label{eq:purification eta p2}
\end{equation}
$F'$ is the fidelity obtained on purifying two entangled pairs of fidelity $F$. $p_2$ is the two-qubit gate reliability parameter, indicating the probability that the CNOT gate performs reliably. Concretely, they modelled the CNOT operation as:
\begin{equation}
    O_{12}\rho = p_2 O_{12}^{\text{ideal}} \rho + \frac{1-p_2}{4} \text{tr}_{12}\{\rho\} \otimes I_{12}
    \label{Eq:p2}
\end{equation}
Where $O_{12}^{\text{ideal}}$ is the ideal CNOT operation, which happens with probability $p_2$. Otherwise, the state is mapped to a completely mixed state. This happens with probability $1-p_2$.

\section{Purification map}
\label{sec:purification map}
The purification map from the Bennett protocol \cite{bennett1996purification}, for two entanglement pairs of fidelities $F$, and no experimental errors is given by (\ref{eq:purification ideal}). The numerator is the probability that the purified state is the desired $\Phi^+$ bell state, conditioned on the event that the purification attempt was accepted, i.e.~the measurements on the target qubits gave the same result. This acceptance probability is given by the denominator. Therefore, if $A$ is the event where the purification yields $\Phi^+$, and $B$, where the purification is accepted, then the fidelity $F'$ after purification is given by:
\begin{equation}
    F' = \frac{P(A|B)}{P(B)} = \frac{F^2 + \frac{(1 - F)^2}{9}}{F^2 + 2F\frac{(1 - F)}{3} + 5\frac{(1 - F)^2}{9}}
\end{equation}
This is easy to visualise from Table \ref{table:probabilities}, where the probabilities of all initial and final state combinations are given for the error-free case. The idea remains the same in the presence of errors. The read-out efficiency $\eta$ is accounted for by considering probabilities that the target qubits might flip from the measurement towards the end of the purification. Similarly, the gate error $\epsilon_g$ introduces a Pauli error right before the measurement, according to error model (\ref{eq:error model_after}). 

\begin{table}[h!]
\centering
\begin{tabular}{|c|c|c|c|c|}
\hline
\multicolumn{2}{|c|}{Initial state} & \multicolumn{2}{c|}{Final state} & \multirow{2}{*}{Probability} \\ \cline{1-4}
Sources & Targets & Sources & Targets & \\ \hline
$\Phi^{\pm}$ & $\Phi^{+}$ & $\Phi^{\pm}$ & $\Phi^{+}$ & $F^2, F(1-F)/3$ \\ \hline
$\Phi^{\pm}$ & $\Phi^{-}$ & $\Phi^{\mp}$ & $\Phi^{-}$ & $F(1-F)/3, (1-F)^2/9$\\ \hline
$\Psi^{\pm}$ & $\Psi^{+}$ & $\Psi^{\pm}$ & $\Phi^{+}$ & $(1-F)^2/9$\\ \hline
$\Psi^{\pm}$ & $\Psi^{-}$ & $\Psi^{\mp}$ & $\Phi^{-}$ & $(1-F)^2/9$\\ \hline
$\Phi^{\pm}$ & $\Psi^{+}$ & $\Phi^{\pm}$ & $\Psi^{+}$ & $F(1-F)/3, (1-F)^2/9$\\ \hline
$\Phi^{\pm}$ & $\Psi^{-}$ & $\Phi^{\mp}$ & $\Psi^{-}$ & $F(1-F)/3, (1-F)^2/9$\\ \hline
$\Psi^{\pm}$ & $\Phi^{+}$ & $\Psi^{\pm}$ & $\Psi^{+}$ & $F(1-F)/3$\\ \hline
$\Psi^{\pm}$ & $\Phi^{-}$ & $\Psi^{\mp}$ & $\Psi^{-}$ & $(1-F)^2/9$\\ \hline
\end{tabular}
\caption{Action of the CNOT gate on initial pairs of fidelity $F$. The probability column indicates the probabilities of the initial states. The first and second probabilities are for the states corresponding to the $+$ and $-$ superscripts in the source column of the initial states respectively. If there is only one entry, then both probabilities are the same.}
\label{table:probabilities}
\end{table}

\section{Full integrals}
\label{sec:full integrals}
In Sec.~{\ref{sec:exponent}}, we introduced a non-recursive estimation of the resource scaling exponent $\Tilde{\lambda}$ (\ref{eq:approx exp}), which involved non-recursive estimations of the average fidelity improvement $g$ (\ref{eq:approx g}) and the product average $\Tilde{P_F}$ of the purification acceptance probabilities (\ref{eq:avg pf}).

We recall that the non-recursive exponent was given by
\begin{align}
    \Tilde{\lambda} &= \log_2 \Tilde{B} + 1\\
    \Tilde{B} &= \left( \frac{2}{P_s \Tilde{P_F}} \right)^{\Tilde{m}},
\end{align}
where
\begin{align}
    \Tilde{m} &= \left\lceil \frac{\Delta F}{g} \right\rceil\label{eq:m_expr}\\
    \Tilde{P_F} &= \lim_{m \rightarrow \infty} \left( \prod_{i = 1}^m P_{F_i} \right)^{1/m} = \exp\left(\frac{\int_{F_0}^{F_t} \log P_FdF}{F_t-F_0}\right).\label{eq:PF_expr}
\end{align}
Using the average fidelity gain 
\begin{equation}
    g = \frac{\int_{F_0}^{F_t} [F'(F) - F] dF}{F_T-F_0},
\end{equation}
we can approximate \eqref{eq:m_expr} analytically by the fraction $\Tilde m \approx \left\lceil\xi/\zeta \right \rceil$ with the numerator 
\begin{equation}
\begin{split}
    \xi=\left(- 16 F_{0} + 16 F_{t}\right)& \left(- F_{0} + F_{t}\right) \left(2 \epsilon_{r} - 1\right)^{4} \\
    &\cdot \left(64 \epsilon_{r}^{6} - 192 \epsilon_{r}^{5} + 240 \epsilon_{r}^{4} - 160 \epsilon_{r}^{3} + 60 \epsilon_{r}^{2} - 12 \epsilon_{r} + 1\right)
\end{split}
\label{eq:m numer}
\end{equation}
and the denominator
\begin{equation}
    \begin{split}
        \zeta = &\left(- 4 F_{0} + 4 F_{t}\right) \left(2 \epsilon_{r} - 1\right)^{2} \cdot \left(64 \epsilon_{r}^{6} - 192 \epsilon_{r}^{5} + 240 \epsilon_{r}^{4} - 160 \epsilon_{r}^{3} + 60 \epsilon_{r}^{2} - 12 \epsilon_{r} + 1\right) \\ 
        &\quad \cdot\bigg(2 \epsilon_{g}^{4} \epsilon_{r}^{2} - 2 \epsilon_{g}^{4} \epsilon_{r} + \epsilon_{g}^{4} - 8 \epsilon_{g}^{3} \epsilon_{r}^{2} + 8 \epsilon_{g}^{3} \epsilon_{r} - 4 \epsilon_{g}^{3} + 22 \epsilon_{g}^{2} \epsilon_{r}^{2} - 22 \epsilon_{g}^{2} \epsilon_{r} + 10 \epsilon_{g}^{2} \\
        &\quad \quad - 28 \epsilon_{g} \epsilon_{r}^{2} + 28 \epsilon_{g} \epsilon_{r} - 12 \epsilon_{g} + 12 \epsilon_{r}^{2} - 12 \epsilon_{r} - \left(2 F_{0} + 2 F_{t}\right) \left(2 \epsilon_{r} - 1\right)^{2} + 5\bigg) \\
        &+ 3 \left(\epsilon_{g} - 1\right)^{2} \left(2 \epsilon_{r} - 1\right)^{2} \cdot \left(2 \epsilon_{r}^{2} - 2 \epsilon_{r} + 1\right) \\
        &\quad\cdot\Bigg[- 2 \left(2 \epsilon_{r} - 1\right)^{5} \left(2 \epsilon_{g}^{2} \epsilon_{r}^{2} - 2 \epsilon_{g}^{2} \epsilon_{r} + \epsilon_{g}^{2} - 4 \epsilon_{g} \epsilon_{r}^{2} + 4 \epsilon_{g} \epsilon_{r} - 2 \epsilon_{g} + 2\right)\\
        &\quad\quad \cdot\operatorname{atan}{\left(\frac{1.5}{4 F_{0} \epsilon_{r} - 2 F_{0} - \epsilon_{r} + 0.5} \right)} \\
        &\quad+ 2 \left(2 \epsilon_{r} - 1\right)^{5} \cdot \left(2 \epsilon_{g}^{2} \epsilon_{r}^{2} - 2 \epsilon_{g}^{2} \epsilon_{r} + \epsilon_{g}^{2} - 4 \epsilon_{g} \epsilon_{r}^{2} + 4 \epsilon_{g} \epsilon_{r} - 2 \epsilon_{g} + 2\right) \\
        &\quad\quad \cdot\operatorname{atan}{\left(\frac{1.5}{4 F_{t} \epsilon_{r} - 2 F_{t} - \epsilon_{r} + 0.5} \right)} \\
        & \quad-\left(- 64.0 \epsilon_{r}^{6} + 192.0 \epsilon_{r}^{5} - 240.0 \epsilon_{r}^{4} + 160.0 \epsilon_{r}^{3} - 60.0 \epsilon_{r}^{2} + 12.0 \epsilon_{r} - 1.0\right) \\
        &\quad\quad \cdot\log{\left(\left(F_{t} - 0.25\right)^{2} + \frac{2.25}{\left(4 \epsilon_{r} - 2\right)^{2}} \right)} \\
        & \quad+\left(- 64 \epsilon_{r}^{6} + 192 \epsilon_{r}^{5} - 240 \epsilon_{r}^{4} + 160 \epsilon_{r}^{3} - 60 \epsilon_{r}^{2} + 12 \epsilon_{r} - 1\right) \\
        &\quad\quad \cdot\log{\left(\left(F_{0} - 0.25\right)^{2} + \frac{2.25}{\left(4 \epsilon_{r} - 2\right)^{2}} \right)}\Bigg]
    \end{split}
    \label{eq:m denom}
\end{equation}
The integral in the expression \eqref{eq:PF_expr} for the average purification success probability $\Tilde{P_F}$ is given by:

\begin{equation}
\begin{split}
    \int_{F_0}^{F_t} \log P_FdF = &
    \sum_\alpha s_\alpha  (- F_\alpha) \log\Bigg[\frac{8 \epsilon_{r} \left(F_\alpha - 1\right) \left(2 F_\alpha + 1\right) \left(\epsilon_{r} - 1\right)}{9} \\
    &\hspace{100pt}+\frac{\left(\epsilon_{r}^{2} + \left(\epsilon_{r} - 1\right)^{2}\right) \left(3 F_\alpha^{2} - 2 F_\alpha \left(F_\alpha - 1\right) + \frac{5 \left(F_\alpha - 1\right)^{2}}{3}\right)}{3} \Bigg] \\
    &\hspace{25pt}+ s_\alpha \Bigg[\frac{2  \left(3 - 6 \epsilon_{r}\right) \operatorname{atan}{\left(\frac{3 - 6 \epsilon_{r}}{\left(F_\alpha - \frac{1}{4}\right) \left(16 \epsilon_{r}^{2} - 16 \epsilon_{r} + 4\right)} \right)}}{16 \epsilon_{r}^{2} - 16 \epsilon_{r} + 4} \\
    &\hspace{60pt}+\frac{1}{4} \log{\left(\frac{\left(3 - 6 \epsilon_{r}\right)^{2}}{\left(16 \epsilon_{r}^{2} - 16 \epsilon_{r} + 4\right)^{2}} + \left(F_\alpha - \frac{1}{4}\right)^{2} \right)}\Bigg] \\
    &- 2 \left(F_t-F_0\right),
\end{split}
\label{eq:pf explicit}
\end{equation}
where $\alpha =0,t$ and the signs $s_0=1$ and $s_t=-1$.

\section{Derivation of the optimal target fidelity}
\label{sec:opt ft}
Here we derive the optimal target fidelity  $F_{t}^{*}$ described in \eqref{eq:opt ft} which minimizes the entanglement resources needed to maintain it across a repeater chain (see Sec.~\ref{sec:exponent}). 

Define the gain $G$ in fidelity $F$ from one purification round as: 
\begin{equation}
    G = F'(F)-F =: \Delta F
    \label{eq:g}
\end{equation}
$G$ in (\ref{eq:g}) is defined for a single purification step, $\Delta m = 1$. Defining a continuous approximation of $G$ that permits a non-integer number of steps: 
\begin{equation}
G \approx \frac{dF}{dm}
\end{equation}
Therefore, the total number of steps $m$ required to go from a starting fidelity $F_0$ to a target fidelity $F_t$:
\begin{equation}
    m \approx \int_{F_0}^{F_t} \frac{dm}{dF} dF = \int dF \frac{1}{G}
\end{equation}
Recall, the total number $B$ (\ref{eq:B}) of entanglement pairs needed in each nesting level:
$$B = \prod_{i=1}^m\frac{2}{P_s P_{F_i}}$$

So, 
\begin{equation}
    \log B = \sum_{i=1}^m \log \frac{2}{P_s P_{F_i}} \approx \int_{0}^{m} \log \frac{2}{P_s P_{F(m')}} dm' = \int_{F_0}^{F_t} \log \frac{2}{P_s P_{F}} \frac{dm}{dF} dF = \int_{F_0}^{F_t} \frac{1}{G} \log \frac{2}{P_s P_{F}} dF
    \label{eq:logB}
\end{equation}
The goal is to minimize (\ref{eq:logB}) with respect to $F_t$. This requires an extremization:
\begin{equation}
    \frac{d(\log B)}{d F_t} = 0
    \label{eq: B extremisation}
\end{equation}
From (\ref{eq:logB}), (\ref{eq: B extremisation}) can be written as: 
\begin{equation}
    \left[\frac{1}{G} \log \frac{2}{P_s P_{F}}\right]_{F = F_t} - \left[\frac{1}{G} \log \frac{2}{P_s P_{F}}\right]_{F = F_0} \frac{dF_0}{dF_t} = 0
    \label{eq:B extremisation explicit}
\end{equation}
Here $F_0$ is taken to be the fidelity obtained on swapping two links of fidelities $F_t$, according to \eqref{Eq:fidelity us} and approximated to $F_0 \approx 2F_t - 1$. On inserting $G$ from (\ref{eq:g}) into (\ref{eq:B extremisation explicit}) and linearising with respect to all errors up to quadratic order, we solve it with respect to $F_t$, and pick the smaller solution:
\begin{equation}
\begin{matrix}
F_t^*\approx\frac{8.31 \epsilon_{g} \epsilon_{r} - 0.40 \epsilon_{g} - 3.35 \epsilon_{r} - 2 \sqrt{0.04 \epsilon_{g}^{2} \epsilon_{r}^{2} + \epsilon_{g}^{2} \epsilon_{r} + 0.45 \epsilon_{g}^{2} - 0.04 \epsilon_{g} \epsilon_{r}^{2} - 0.4 \epsilon_{g} \epsilon_{r} + 0.07 \epsilon_{g} + 0.007 \epsilon_{r}^{2} - 0.001 \epsilon_{r}} + 0.61}{8.36 \epsilon_{g} \epsilon_{r} + 0.8 \epsilon_{g} - 3.37 \epsilon_{r} + 0.61} \\
\approx\frac{-1.16\epsilon_g - 4.28\sqrt{\epsilon_g^2 + 0.15\epsilon_g} + 1.9}{2.66\epsilon_g + 1.9}
\end{matrix}
\label{eq:opt ft full}
\end{equation}
\nocite{*}

\section{CNOT-based purification}
\label{sec:bennett}
\begin{figure}
    \centering
    \begin{subfigure}[b]{0.8\textwidth}
        \centering
        \includegraphics[width=\textwidth]{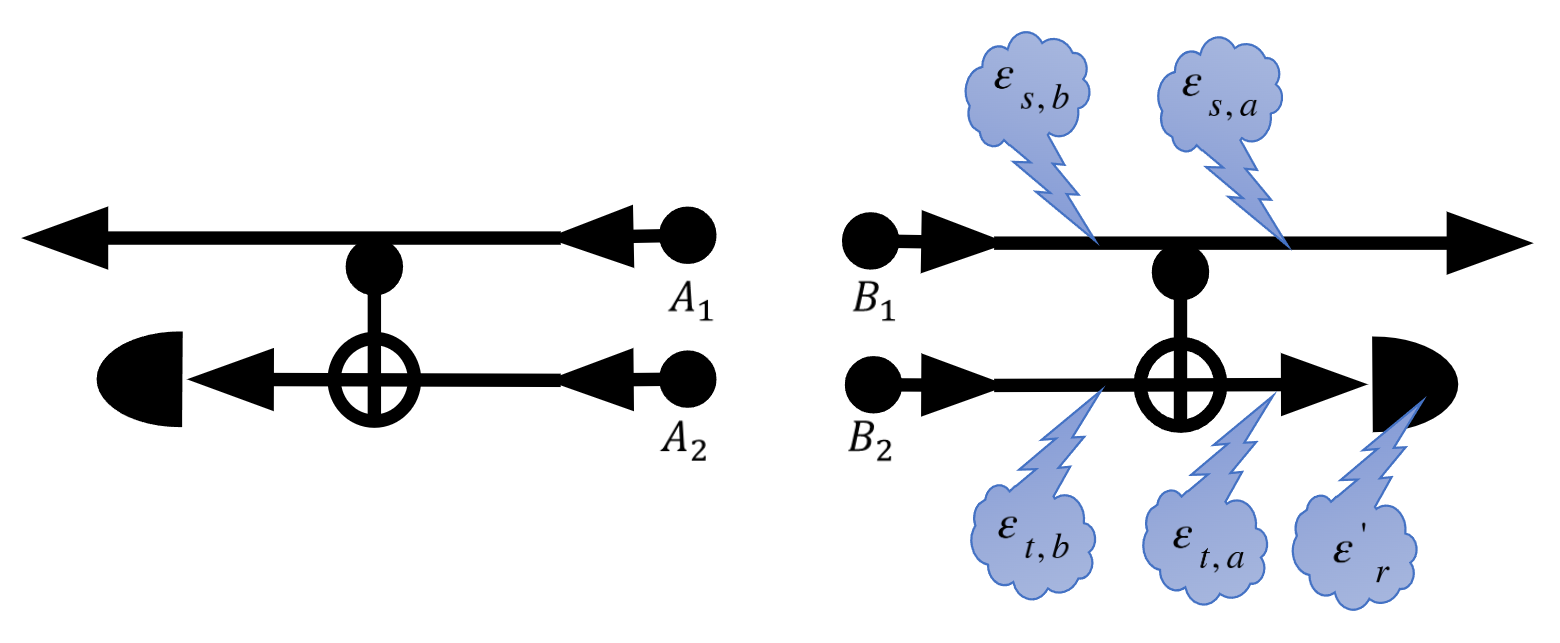}
        \caption{}
        \label{fig:bennett purification}
    \end{subfigure}
    \hfill
    \begin{subfigure}[b]{0.8\textwidth}
        \centering
        \includegraphics[width=\textwidth]{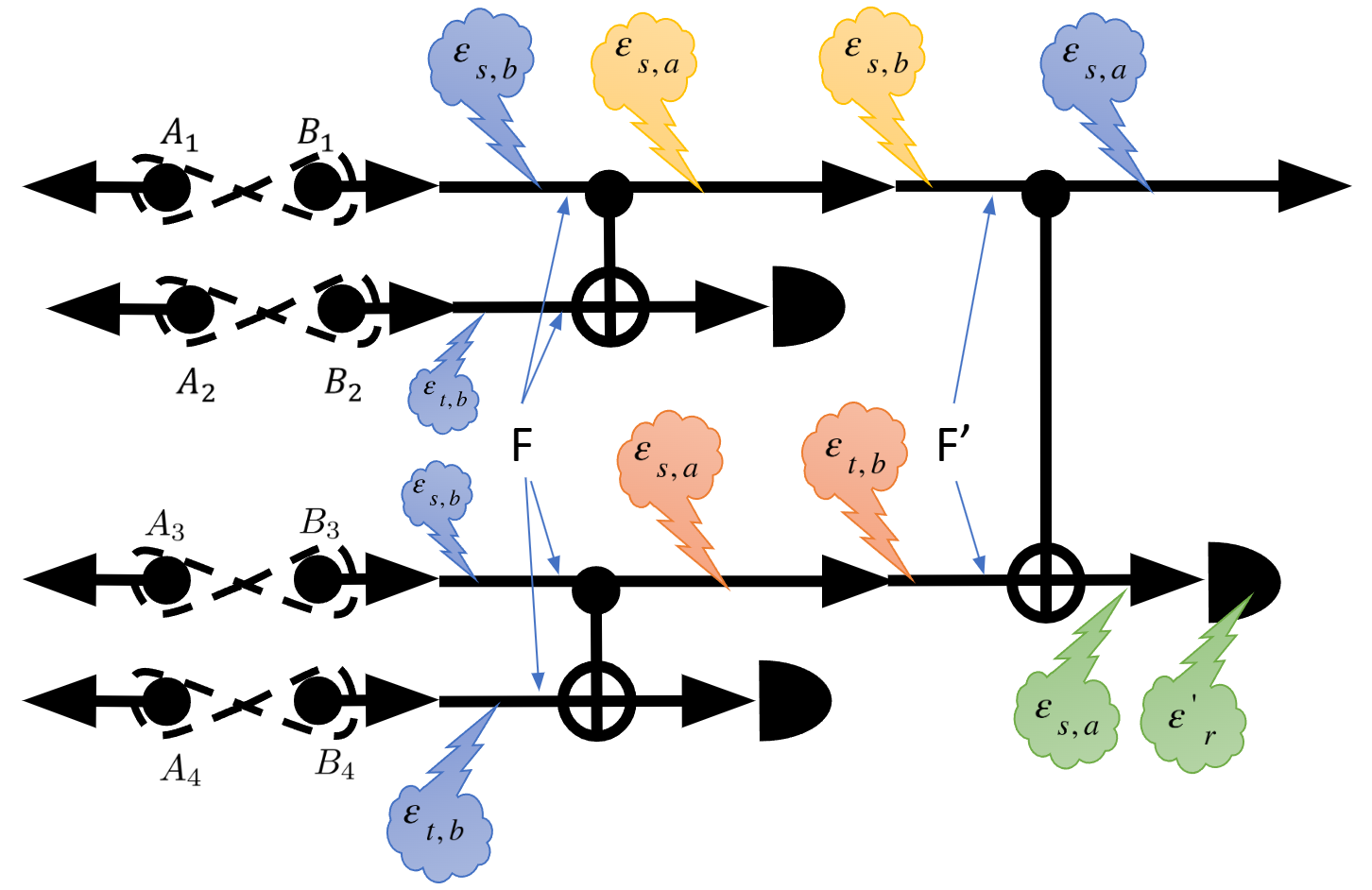}
        \caption{}
        \label{fig:bennett 2}
    \end{subfigure}
    \caption{Purification: (a) \(A_1B_1\) and \(A_2B_2\) approximate the same entangled state, e.g.\ $\left|\Phi^+\right\rangle\propto\left|00\right\rangle+\left|11\right\rangle$ with fidelity \(F\). CNOT gates are applied between \(A_1 A_2\) and \(B_1 B_2\) respectively, and control bits \(A_1B_1\) are kept, if measurements of the target bits \(A_2\) and \(B_2\) in the computational basis yield the same results, and are discarded otherwise. $\epsilon_{c/t, b/a}$ are gate errors on the control/target qubits, that occur before/after the CNOT operation. $\epsilon_r$ is the read-out error on the target qubit. (b) when multiple rounds of purification are performed, errors can be grouped together (yellow, orange and green) to form effective errors. For details see text.}
    \label{fig:combined_bennett}
\end{figure}

We illustrate the purification circuit outlined in Sec.~\ref{sec:related work}, in Fig.~\ref{fig:bennett purification}. The CNOT-based schemes involve two pairs of qubits, \(A_1B_1\) and \(A_2B_2\), which approximate the same entangled state, e.g.\ $\left|\Phi^+\right\rangle\propto\left|00\right\rangle+\left|11\right\rangle$ with fidelity \(F\). CNOT gates are applied between \(A_1 A_2\) and \(B_1 B_2\) respectively, and control bits \(A_1B_1\) are kept, if measurements of the target bits \(A_2\) and \(B_2\) in the computational basis yield the same results, and are discarded otherwise. Successful purification reduces bit-flip errors in the state, yielding a higher fidelity state. 

As explained in Sec.~\ref{sec:purification}, the errors $\epsilon_{s,b}$ ($\epsilon_{s,a}$) act on the control qubit before (after) the ideal interaction, whereas $\epsilon_{t,b}$ ($\epsilon_{t,a}$) act on the target qubit before (after) the interaction. The physical error $\epsilon'_r$ affects the read-out of the target qubit, see Fig. \ref{fig:bennett purification}.

The read-out error can be modeled as a misidentification of the state, by using the measurement operators
\begin{equation}
\begin{aligned}
    P_0^{\eta_0'} &= \eta_{0}' |0\rangle\langle0| + \epsilon_0'|1\rangle\langle1| \\
    P_1^{\eta_1'} &= \eta_1' |1\rangle\langle1| + \epsilon_1'|0\rangle\langle0|
\end{aligned}
\label{Eq:measure_SI}
\end{equation}
to describe the imperfect measurement of a single qubit in state $|0\rangle$ ($P_0^{\eta_0'}$) or $|1\rangle$ ($P_1^{\eta_1'}$).
In a deterministic measurement, $\eta_0' = 1-\epsilon_1'$ and $\eta_1' = 1-\epsilon_0'$. For the sake of simplicity, we assume a deterministic and symmetric read-out error, i.e. \begin{equation}\eta'=1-\epsilon_r' = \eta_0'=\eta_1'=1-\epsilon_0'=1-\epsilon_1'\end{equation}

Therefore, if the qubit to be measured is $\rho = |0\rangle\langle0|$, then the expectation values are $\langle P_0^{\eta'} \rangle = \eta'$ and $\langle P_1^{\eta'} \rangle = 1-\eta'$, meaning that there is a probability $\eta'$ that the measurement (\ref{Eq:measure}) gives the correct result $|0 \rangle \langle 0|$. 

Finding a general model for the gate error, however is more complicated: different experimental platforms implement entanglement gates using vastly different methods, such that a platform-independent error model necessarily requires some simplifications to produce understandable predictions. Here, we restrict our analysis to incoherent errors, assuming coherent errors, such as over-rotations can be suppressed well enough by calibration of analog control pulses, or compensated using decoupling or robust pulse sequences. For incoherent errors, i.e.\ errors that occur independent of each other, we use a Lindblad master equation to describe the open quantum system. The Lindbladian superoperator
\begin{equation}
    \mathcal{D} \left[s\right] \rho = s \rho s^\dagger - \frac 1 2 \left(ss^\dagger \rho +  \rho ss^\dagger \right)
\end{equation}
describes the incoherent action of an operator $s$ onto the system, described by the density operator $\rho$. For an idealized system Hamiltonian $H_\textrm{sys}$, the interaction with a bath is modeled by
$$\dot{\rho} = \mathcal L\rho = -\frac i \hbar \left[H_\textrm{sys}, \rho\right] + \frac{\gamma_z}{2} \mathcal D[\sigma_z] + \kappa (n_{th}+1) \mathcal D[\sigma^-] + \kappa n_{th} \mathcal D[\sigma^+].$$
Here, $\gamma_z$ is the pure dephasing rate, $\kappa$ describes the coupling to a thermal bath with a thermal occupation number $n_{th}$.
In the limit of a hot bath, i.e.\ the occupation number $n_{th}$ of the bath coupled to the qubit is large, such that $n_{th} \approx n_{th}+1$, the raising ($\sigma^+$) and lowering operator ($\sigma^-$) have a similar effect on the system. We can thus approximate the coupling of the system to the environment as
$$\dot{\rho} = -\frac i \hbar \left[H_\textrm{sys}, \rho\right] + \frac{\gamma_z}{2} \mathcal D[\sigma_z] + \frac{\gamma_x}{2} \mathcal D[\sigma_x] + \frac{\gamma_y}{2} \mathcal D[\sigma_y]$$ 
for two entangled qubits with equal expectation value for the excitation number $\left\langle \sigma^+\sigma^-\right\rangle=1/2$, introducing the bit flip rates $\gamma_x$ and $\gamma_y$. 

Without specific assumptions about the system Hamiltonian $H_\textrm{sys}$, we cannot make specific predictions about the behavior of the system, with numerous types of errors being possible.  A common approach is to assume an instant, ideal gate, and error happening before and after this gate. This corresponds to phase ($\gamma_z$) and bit flip ($\gamma_x$, $\gamma_y$) Pauli errors. This means, a Pauli operator $\sigma_j^{i}$ ($j=x,y,z$ and $i=s,t$) acting on the source ($s$, e.g.\ $B_1$ in Fig.~\ref{fig:bennett purification}) or target qubit ($t$, e.g.\ $B_2$ in Fig.~\ref{fig:bennett purification}) is applied to the density matrix $\rho$ of the system with a certain probability $p_{i,j, k}$. This can happen either before ($k=b$) or after ($k=a$) the ideal CNOT gate is applied.  We summarize the probability for any error to happen to the source or target qubit before or after the idealized gate by the probabilities (see Fig.~\ref{fig:bennett purification})
\begin{equation}
    \epsilon_{i,k}=\sum_j  p_{i,j, k}.
    \label{eq:error_probability}
\end{equation}

Another common type of error occurs, for example, when the source qubit flips during the gate execution, resulting in an unconditional rotation of the target qubit. Hence, this corresponds to executing a different gate instead of the CNOT gate, and cannot simply be described by Pauli errors. Here, however, the target qubit is measured directly after the gate operation, such that this failed-gate error can be captured to first order by adapting  $p_{s,j, b}$ (for an unconditional rotation of the target qubit) and $p_{s,j, a}$ (for non-executed gates) accordingly. 

Hence, in the case of CNOT-based purification, a Pauli-error model captures the most important error types during gate operation. We note that this error model differs from the one used in Ref.~\cite{briegel1998quantum, dur1999quantum, hartmann2007role}, which assumed a projection onto a mixed state, whenever an error occurs. While this error model is effectively the same as a Pauli-error model when analyzing the creation of entangled states, it can produce vastly different results in the context of entanglement purification: Projection onto a mixed state implies that the error flips the target qubit in 50\% of the cases, leading to a detection that an error occurred. In contrast, Pauli errors happening on the source qubit after the gate execution, corresponding to $p_{s,j, a}$, are never detected, as they do not influence the target qubit. 

In order to further reduce the complexity of the model, we group the errors \eqref{eq:error_probability} which occur right after each other, see also Fig.~\ref{fig:bennett 2}.
The effective gate-errors $\epsilon_{g,1} = \epsilon_{s,a}+\epsilon_{s,b}$ (Fig. \ref{fig:bennett 2}, yellow) and $\epsilon_{g,3} = \epsilon_{s,a}+\epsilon_{t,b}$ (Fig.~\ref{fig:bennett 2}, red), effectively only act on the respective control qubits $B_1$ and $B_3$ of the first layer of the entanglement protocol. They appear after the interaction, i.e.\ they will not influence the target qubits $B_{2,4}$ and, hence, conditioning of the purified state, but effectively add to its error. 
To simplify the analysis, we assume a symmetrized effective gate error 
\begin{equation}\epsilon_{g}\approx\epsilon_{g,1}/2+\epsilon_{g,3}/2 = \epsilon_{s,a} +\epsilon_{s,b}/2+\epsilon_{t,b}/2 \end{equation}
We note that $\epsilon_{s,a}$ has the biggest effect on this effective gate error. 
Here, we effectively assume, that the role of the qubits is interchangeable, i.e.\ no swap errors are taken into account. These could be added to the model in first order by adapting $\epsilon_{t,b}$. 

In most experimental platforms, the errors are not equally weighted, with dephasing often playing a dominant role. To take this into account, we can decompose the effective gate error into the individual Pauli operator contributions, i.e.\ the conditional probability for an error that occurred to be described by a Pauli operator along a certain axis $j$: $p_j=$ P(effective $\sigma_j$ error occurred$|$gate error occurred) (for $j=x,y,z$)
\begin{equation}
    p_j \approx \frac{p_{s,j, A} + \frac{p_{s,j, B}}{2} + \frac{p_{t,j, B}}{2}}{\epsilon_g}.
\end{equation}
We note that $x$ and $y$ errors have the same effect on the purification map \eqref{eq:purification eta eg}, such that we can summarize $\epsilon_g(p_x+p_y)$ as bit-flip-errors and $\epsilon_g p_z$ as phase flip errors.

For the effective readout error, we sum up bit-flip errors on the target qubit after the interaction $p_{t,x,a}$ and $p_{t,y,a}$ with the physical readout errors $\epsilon_r'$ to obtain an effective readout error (see also Fig.~\ref{fig:bennett 2})
\begin{equation}
    \epsilon_r=1-\eta=\epsilon_r'+p_{t,x,a}+p_{t,y,a}.
\end{equation}
With the effective gate error $\epsilon_g$ and read-out error $\epsilon_r$, we can now describe the effect of entanglement purification on the state fidelity at the moments before the idealized gates by \eqref{eq:purification eta eg}. This is indicated in Fig.~\ref{fig:bennett 2} by the fidelity $F$ before, and $F'$ after the purification. We note, that these points are not necessarily accessible experimentally: when purified entanglement is characterized, this happens before the next purification round starts, i.e.\ before the errors $\epsilon_{i,b}$ (for $i=s,t$) happen. However, when the purified entanglement is used, e.g.\ for entanglement swapping or teleportation, another gate needs to be applied, such that $F'$ can be understood as representing the usable fidelity of the state, in an applied network scenario. 

In particular, this simplifies the error treatment of the entanglement swapping step in the first generation repeater protocol:
neglecting single qubit control errors, the effective readout error for the two qubits are
$\epsilon_{r,S}^{t} = \epsilon_r$ and $\epsilon_{r,S}^{s} = \epsilon_r'+p_{s,x,a}+p_{s,z,a}$ (or, depending on the implementation of the Bell state measurement, $\epsilon_{r,S}^{s} = \epsilon_r'+p_{s,y,a}+p_{s,z,a}$).
For the sake of simplicity, we again assume a symmetrized error $\epsilon_{r,S}=1-\eta_S = \epsilon_{r,S}^{t} =\epsilon_{r,S}^{s} = \epsilon_r$. Thus, entanglement swapping can be modeled as an ideal two qubit gate, followed by two measurements with effective error $\epsilon_r$ each. We note that in this case, the gate error is to first order irrelevant, as it was already taken into consideration in the previous purification step, see eq.~\eqref{Eq:fidelity us}. 

Finally, we need to connect our theoretical model with experimentally observable parameters. While a tomography of the gate would allow to estimate all $p_{i,j, k}$, this requires long measurements, and is rarely done. Instead, most systems merely report the entanglement fidelity between two qubits, respectively measurements that allow for extracting this fidelity. Using our error model, a typical entanglement experiment will result in a linearized entanglement fidelity  $F_\textrm{ent}\sim 1-\epsilon_{s,a}-\epsilon_{s,b}-\epsilon_{t,a}-\epsilon_{t,b}$, after correcting for state preparation and measurement errors. 
Assuming all four errors contribute equally, we thus find $\epsilon_{g}\sim \frac{1-F_{ent}}{2}$ (see also Appendix \ref{sec: table assumptions}). Hence, it can be interpreted as the average probability of an error affecting one of the two qubits in an entanglement operation.

\section{Assumptions in Table \ref{tab:lambdas}}
\label{sec: table assumptions}
Since we consider the effective gate error $\epsilon_g$ only on the control qubit in the purification protocol, we take it to be $\frac{1-F_2}{2}$, where $F_2$ is the unidirectional two-qubit gate fidelity, corrected for state preparation and measurement errors, reported in literature. Although we always absorb $\epsilon_{t,a}$ into $\epsilon_r$ for the target qubit, we take $\epsilon_r$ to be $1-\eta$, where $\eta$ is the reported read-out/measurement fidelity, creating the most optimistic scenario from the reported values. The exponent is computed by taking the starting fidelity to be the optimal one, $F_t^*$ (\ref{eq:opt ft}) given the error combination $(\epsilon_r, \epsilon_g)$. In an experiment, this could mean that extra purification steps are required before starting the quantum repeater protocol, if optical entanglement of remote repeaters has a lower fidelity than $F^*_T$. This overhead in not taken into account here, as it does not effect the scaling. Furthermore, we do not account for the decoherence experienced by the qubits whilst waiting for the classical messages in the purification and swapping procedures, for the calculation of the exponent. This would effectively add to the gate error, such that the exponents $\lambda$ ought to be considered as lower bounds to the actual resource scaling of the considered quantum network.
Additionally, linear prefactors due to photon collection efficiency, and quantum frequency conversion to the telecommunication wavelength range are not considered here in the computation of the exponent. Although these have been demonstrated for all platforms described above, achieving high fidelity and efficiency remains a challenge for most platforms (see below).
When computing the maximum number links $D^*$, we use literature values for coherence time and entanglement rate. The table indicates which experiments did perform quantum frequency conversion when demonstrating the reported entanglement rates. 

The errors $\epsilon_g$ and $\epsilon_r$, and the entanglement rate $R$ for any platform were not taken from the same experimental setup, therefore the provided values may not be possible to attain concurrently in a single setup.

In particular, a full demonstration of optical entanglement of superconducting qubits has, to our knowledge, not yet been demonstrated. We assume that all individual components of a mechanically mediated optical interface could be implemented while maintaining errors and rates reported in literature, i.e.\ optical entanglement of two nanomechanical resonators, and piezoelectric transduction to lumped element superconducting qubits of the highest coherence time. Other methods of generating remote entanglement of superconducting qubits across a room-temperature channel may prove more efficient, but this setup can provide an outlook on the possibilities of optically entangled superconducting qubits. 

\bibliography{apssamp}

\end{document}